\documentclass[twocolumn,superscriptaddress,preprintnumbers,prl]{revtex4}

\usepackage{ORI_Group_style}
\newcommand{\uv}{\boldsymbol{e}}
\newcommand{\cyl}{(\rho,z,\w)}

\begin{document}

\title{Ultrafocused Electromagnetic Field Pulses with a Hollow Cylindrical Waveguide}

\author{P.~Maurer}
    
\affiliation{Institute for Quantum Optics and Quantum Information of the
Austrian Academy of Sciences, A-6020 Innsbruck, Austria.}

\affiliation{Institute for Theoretical Physics, University of Innsbruck, A-6020 Innsbruck, Austria.}

\author{J.~Prat-Camps}
    
\affiliation{Institute for Quantum Optics and Quantum Information of the
Austrian Academy of Sciences, A-6020 Innsbruck, Austria.}

\affiliation{Institute for Theoretical Physics, University of Innsbruck, A-6020 Innsbruck, Austria.}

\author{J.~I.~Cirac}

\affiliation{Max-Planck-Institut f\"ur Quantenoptik,
Hans-Kopfermann-Strasse 1, D-85748, Garching, Germany}

\author{T.~W.~H\"ansch}

\affiliation{Ludwig-Maximilians-Universit\"at M\"unchen, Fakult\"at f\"ur Physik, Schellingstrasse 4/III, 80799 M\"unchen, Germany}

\affiliation{Max-Planck-Institut f\"ur Quantenoptik,
Hans-Kopfermann-Strasse 1, D-85748, Garching, Germany}

\author{O.~Romero-Isart}
    
\affiliation{Institute for Quantum Optics and Quantum Information of the
Austrian Academy of Sciences, A-6020 Innsbruck, Austria.}

\affiliation{Institute for Theoretical Physics, University of Innsbruck, A-6020 Innsbruck, Austria.}
    
\begin{abstract}

We theoretically show that an externally driven dipole placed inside a cylindrical hollow waveguide can generate a train of ultrashort and ultrafocused electromagnetic pulses. The waveguide encloses vacuum with perfect electric conducting walls. A dipole driven by a single short pulse, which is properly engineered to exploit the linear spectral filtering of the cylindrical hollow waveguide, excites longitudinal waveguide modes that are coherently re-focused at some particular instances of time. A dipole driven by a pulse with a lower-bounded temporal width can thus generate, in principle, a finite train of arbitrarily short and focused electromagnetic pulses. We numerically show that such ultrafocused pulses persist outside the cylindrical waveguide at distances comparable to its radius. 

\end{abstract}    
    
\maketitle

The precise generation and control of spatiotemporal properties of electromagnetic (EM) fields is an enabling tool for science and technology. Spatial distributions in the far field can be controlled by tailoring the EM properties of media~\cite{Leonhardt2006,Pendry2006}. These so-called metamaterials have been proposed for perfect imaging by either using a negative refractive index to amplify evanescent waves~\cite{Pendry2000}, or an ordinary optical medium whose positive refractive index is continuously varied~\cite{Leonhardt2009}. Regarding temporal properties, the generation of frequency combs, namely a train of coherent ultrashort pulses, has lead to a myriad of applications in precision spectroscopy, optical clocks, astronomical observation, precision ranging, and gas sensing, to mention a few~\cite{Newbury2011,Udem1999,Diddams2001,Steinmetz2008,Schuhler2006,Rieker2014}. Frequency combs are typically generated by using non-linear EM media, such as by mode-locking lasing longitudinal modes by means of a saturable absorber~\cite{Holzwarth2000, Cundiff2003} or by parametric frequency conversion in optical microresonators~\cite{Kippenberg2011}.

\begin{figure}[t]
\begin{center}
\includegraphics[width=0.50\textwidth]{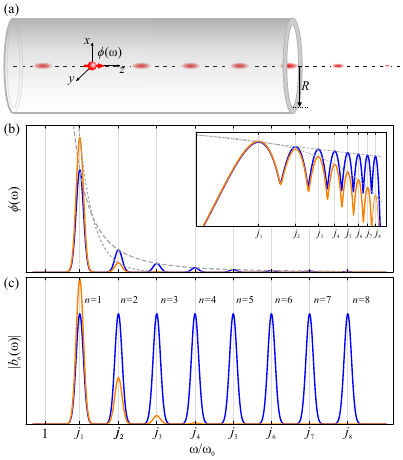}
\caption{(a) Sketch of the setup. The dipole (in red) is shown inside the hollow cylindrical waveguide with inner radius $R$ (gray). Some peaks of electromagnetic energy density are illustrated in red, showing their focusing around the symmetry axis. Outside the cylinder, the main features are preserved up to a distance $\sim R$. (b) Dipole spectra given by \eqnref{eq:FTSpectrum} with $N=8$ and $f(\w) = 1/\sqrt{\w^5}$ (blue line), $f(\w) = \exp[-\w/(1.2\w_0)]$ (orange line). The inset shows plots in a logarithmic scale. (c) Plots of $|b_n(\omega)|$ at $\rho=0$ for the two spectra.}
\label{Fig.1}
\end{center}
\end{figure}

In this Letter, we propose an alternative tool, based neither on spatially structured nor on non-linear EM media, to generate tiny spatiotemporal features of EM fields. In particular, we devise a scheme to generate a train of ultrashort pulses that are ultrafocused in space. This is achieved by placing a driven point dipole inside an otherwise empty hollow cylindrical waveguide of radius $R$ with perfect electric conducting walls, see~\figref{Fig.1}a.  The oscillating dipole has a mean frequency $\bar \w \gtrsim c/R$, where $c$ is the speed of light in vacuum. By properly engineering the temporal dependence of the driven dipole, and in particular, by fine-tuning its spectral tail at high frequencies, we theoretically show that far from the dipole a train of ultrafocused EM pulses is generated.  Each pulse has a temporal width $\sim R/(cN)$ and its EM energy near the axis is mostly carried by the longitudinal components that exhibit a transverse spot size given by $\sim R/N$. Here, $N$ is the number of waveguide modes that the high frequency tail of the driven dipole's spectrum excites. This effect is thus solely based on the geometric properties of the  hollow cylindrical waveguide, which acts as a linear spectral filter, and the fine-tuning of the spectral distribution of the driven dipole. These analytical results are corroborated by numerical simulations. We further show that for a finite waveguide, these ultrashort and ultrafocused EM pulses persist at a distance $\sim R$ outside the cylindrical waveguide. We remark that the discussed mechanism  to generate EM hot-spots is different from superoscillatory fields~\cite{Berry2006,Rogers2013}, which are often realized by illuminating carefully designed optical masks with coherent monochromatic light. These can also focus EM energy in a sub-wavelength spot size far from the source, but only a small fraction of the total energy.

To be more precise, let us consider a point dipole inside an infinite hollow cylinder of radius $R$, whose symmetry axis is aligned with the $z$-axis.  The border of the cylinder, which encloses vacuum, consists of a perfect electric conductor. This sets the boundary condition $\uv_\rho\times\boldsymbol{E}(\rho,z,t)|_{\rho=R}=0$ for the electric field in cylindrical coordinates, where $\rho \equiv \sqrt{x^2+y^2}$ and $\uv_\rho$ is the radial unit vector. The dipole moment is aligned with the symmetry axis, see~\figref{Fig.1}a. Hereafter, we concentrate on a magnetic dipole but the results hold analogously for an electric dipole, see~\cite{SM}. The  magnetic dipole moment is given by $\boldsymbol{m}(t)=m\, \phi(t)\uv_z$, where $m>0$ and $\phi(t)$ is a dimensionless real scalar function  that parametrizes the temporal dependence of $\abs{\boldsymbol{m}(t)}$. The temporal dependence of the point dipole is assumed to be induced by external driving. The magnetization associated to the magnetic dipole, $\boldsymbol{M}(\rho,z,t)\equiv \boldsymbol{m}(t)\delta(\rho)\delta(z)/(2\pi \rho)$, corresponds to a current density given by $\nabla\times\boldsymbol{M}(\rho,z,t)$. Our goal is to derive  the EM field generated by the dipole in the far field $z \gg R$ and analyze its spatiotemporal features.

To this end, we analytically solve Maxwell's equations in the spectral domain. In the following, we sketch the derivation and refer to~\cite{SM} for further details. First, let us define the dipole spectrum $\phi(\w)\equiv\int_{\mathds{R}}\text{d}t\,\phi(t)\exp(\im\w t)$, which (as discussed below) is crucially chosen to be real. We normalize it as $\int_{0}^{\infty}\text{d}\w \phi(\w)=1$. In the framework of vector polarization potentials~\cite{Stratton2007}, the problem is reduced to find the solution of the inhomogeneous scalar Helmholtz equation 
\be
\pare{\grad^2 +k^2}\Psi\cyl=- \frac{\mu_0m \phi(\w) \delta(\rho)\delta(z)}{2\pi \rho},
\ee
with the boundary condition $\partial_\rho\Psi(\rho,z,\w)|_{\rho=R}=0$.  Here, $\mu_0$ is the vacuum permeability and $k^2 \equiv (\w/c)^2$. From the solutions $\Psi\cyl$ of the Helmholtz equation, one can then readily obtain the magnetic field $\boldsymbol{B} \cyl$ and $\boldsymbol{E} \cyl$~\cite{SM}. The symmetry of the problem allows to expand the solutions $\Psi\cyl$ in a complete set of orthogonal functions that fulfill the boundary condition. That is, 
\be
\Psi(\rho,z,\w)=\int_\mathds{R}\text{d}k_z\sum_{n>0}C_n(k_z,\w)J_0 \pare{\frac{j_n \rho}{R}}e^{\im k_z z},
\ee
where $J_m(x)$ denotes the $m$th order Bessel function of the first kind and $j_n$ is the $n$th non-zero root of $J_1(x)$, namely $J_1(j_n)=0$. The expansion coefficients $C_n(k_z,\w)$ can be obtained using the orthogonality relations and the $k_z$ integral can be performed using complex analysis~\cite{SM}. One then analytically obtains that the propagating electric and magnetic fields are given by
\bea
\boldsymbol{B}\cyl &=& B_z\cyl\uv_z+\frac{\im}{\omega} \frac{\partial E_\varphi\cyl}{\partial z}\uv_\rho, \\
\boldsymbol{E}\cyl&=&E_\varphi\cyl\uv_\varphi.
\eea
This corresponds to a magnetic (electric) field with longitudinal and radial (azimuthal) components. These components can be written as an infinite sum of discrete modes
\begin{align} \label{eq:BzModes}
B_z\cyl&=B_0\sum_{n>0}b_{n}(\rho,\w)\delta_n(\omega)\exp[\im \alpha_n(z,\w)], \\
E_\varphi\cyl&=E_0\sum_{n>0}e_{n}(\rho,\w)\delta_n(\omega)\exp[\im \alpha_n(z,\w)],
\end{align}
where $B_0\equiv \mu_0 m/(2\pi R^3)\equiv -E_0/c$. The modes are determined by the phase and amplitude functions
\bea
\alpha_n(z,\w)&\equiv& \text{sign}(\w)\sqrt{\tilde{\w}^2-j_n^2}|\tilde{z}|, \\
\delta_n(\w)&\equiv& \frac{\text{sign}(\w)\Theta(|\tilde{\w}|-j_n)}{\sqrt{j_n}\sqrt{(\tilde{\w}/j_n)^2-1}},\\ \label{eq:bn}
b_{n}(\rho,\w)&\equiv& \im\phi(\w)\sqrt{j_n^3}\frac{J_0(j_n\tilde{\rho})}{J_0^2(j_n)}, \\
e_{n}(\rho,\w)&\equiv& \tilde\w\phi(\w)\sqrt{j_n}\frac{J_1(j_n\tilde{\rho})}{J_0^2(j_n)}.
\eea
Here, $\Theta(x)$ is the Heaviside function and we defined dimensionless frequencies as  $\tilde{\w}\equiv\w/\w_0\equiv\w c/R$ and spatial coordinates as $\tilde{\rho}\equiv\rho/R$ and $\tilde{z}\equiv z/R$.

The analytical solution in the spectral domain unveils some key features of the generated EM fields: (i) the cylinder acts as a high-pass filter since the field components vanish for $|\tilde\w|<j_1$; (ii)  the phase at the resonance frequencies $\tilde \w_n\equiv j_n$ vanishes, namely $\alpha_n(z,\w_n)=0$ (thanks to the fact that $\phi(\w)$ is chosen to be real, namely that the driven dipole has the same spectral phase at all $\w_n$); (iii)  the resonance frequencies are quasi-equidistant, namely $j_n\approx(n+1/4)\pi$~\cite{Olver1974}; (iv) the amplitude $\delta_n(\w)$  diverges at $\omega_n$ as $\text{lim}_{\epsilon\rightarrow 0^+}\delta_n(\omega_n+\epsilon)\propto 1/\sqrt{\epsilon}$, a peak that is equal for all modes; (v) on the $z$-axis ($\rho=0$), the EM field is purely longitudinal and magnetic since $e_{n}(0,\w)=0$; (vi) the peaks of the longitudinal component of the magnetic field at $\w_n$ for $\rho=0$ are given by $b_n(0,\w_n)\approx(\im\pi/2)\phi(\omega_n)\sqrt{j_n^5}$, where $j_nJ_0^2(j_n)\approx 2/\pi$ has been used. 

In summary, the EM field near axis has predominantly a longitudinal magnetic field component  whose spectrum has the form of quasi-equidistant teeth, all with the same phase, but, in general, with different amplitudes. This is the result of the linear spectral filtering of the hollow cylindrical waveguide. According to feature (vi), all modes can equally contribute by tuning the spectrum of the driven dipole such that $\phi(\w_n) \propto 1/\sqrt{\w_n^5}$ for all modes. This leads to $b_n(0,\w_n)=\text{constant}$, such that each tooth in the frequency spectrum has the same height and phase.  This is illustrated in~\figref{Fig.1}(b,c),  where dipole spectra given by $\phi(\w) = g(\w) + g(-\w) $ with
\be \label{eq:FTSpectrum}
g(\w) \equiv \mathcal{N}\sum_{n=1}^N   f(\w_n) \exp \spare{ - \frac{\pare{\w-\w_n}^2}{\sigma^2}}
\ee
($\sigma \ll \omega_0$) are plotted. Here $\mathcal{N}$ is the normalization constant and $f(\w)$ the peak function. An EM field with a spectrum of $N$ quasi-equidistant coherent peaks is thus generated by fine tuning $f(\w) = 1/\sqrt{\w^5}$ [blue line in \figref{Fig.1}(b,c)]. See the difference with the non-tuned spectrum $f(\w) =\exp[-\w/(1.2 \w_0)]$ [orange line in~\figref{Fig.1}(b,c)].  

Therefore, one expects that by engineering a time-dependent dipole such that its spectrum is given by \eqnref{eq:FTSpectrum} with $f(\w) = 1/\sqrt{\w^5}$, one generates an EM field where $B_z(\rho,z,t)$ on the axis consists of a train of pulses, each of duration $\delta t \approx T/N$, temporally separated by $T \sim 2R/c$. Remarkably, each pulse is given by a coherent superposition of the first $N$ zeroth-order Bessel functions, see~\eqnref{eq:BzModes}, which thus should lead to a transverse spot size given by $\delta\rho \approx R/N$~\cite{Maurer2016}.  Off-axis, where $E_\varphi$ is non-zero, one expects these spatiotemporal features to disappear since neither $b_n(\rho,\w_n)$ nor $e_n(\rho,\w_n)$ can be made constant for all $n$. Thus, the energy density $2 u(\rho,z,t) \equiv \epsilon_0 |\boldsymbol{E}(\rho,z,t)|^2+|\boldsymbol{B}(\rho,z,t)|^2/\mu_0$ should show similar spatiotemporal features as those shown by the longitudinal component of the magnetic field. The larger the number $N$ of excited modes is, the more dominant these features are. Note, however, that the mean frequency $\bar \w \equiv \int_0^\infty \text{d} \w \phi(\w) \w$, calculated using \eqnref{eq:FTSpectrum} with $f(\w) = 1/\sqrt{\w^5}$ and $\sigma\ll\omega_0$, can be numerically upper bounded by $\bar \omega /\w_0<8.2$. Hence, $\bar \w$ barely increases with $N$. In essence, the fine-tuned spectrum optimally exploits the linear spectral filtering provided by the hollow cylinder, as given by \eqnref{eq:bn}.

\begin{figure}[t]
\begin{center}
\includegraphics[width=0.50\textwidth]{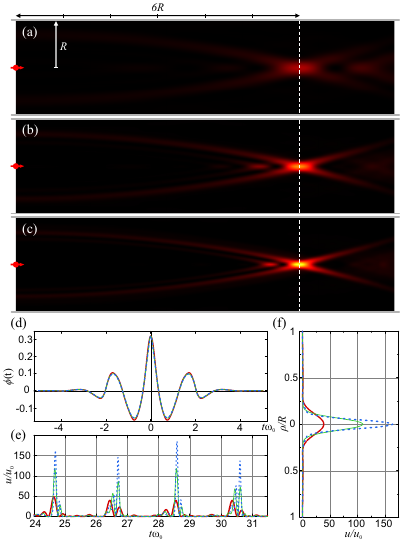}
\caption{Snapshots of normalized EM energy density $u/u_0$, where $u_0\equiv\mu_0 m^2/(4\pi^2 R^6)$, for dipole spectra that excite $N=4,8$, and $12$ modes [(a), (b), and (c) respectively]. Plots are made at similar times, $t \omega_0 \approx 28.5$, corresponding to the third peak of panel (e) [all with the same color scale, ranging from 0 (black) to 230 (white)]. (d) Plot of the temporal dependence of the magnetic moment, $\phi(t)$. (e) Energy density at $\rho=0$, $z=6R$ as a function of time, showing the train of pulses. (f) Energy density distribution at a distance $z=6R$ from the source. Thick-red, thin-green, and dotted-blue lines refer to magnetization spectra with $N=4,8$, and $12$ modes, respectively.}
\label{Fig.2}
\end{center}
\end{figure}

Let us confirm these results by numerically solving, using COMSOL Multiphysics, the Maxwell equations in the temporal domain. We consider a hollow cylinder of a certain length with perfect electric conducting walls. At the two ends, perfectly absorbing transverse walls emulate the infinite length of the cylinder.
Regarding the magnetic dipole, we use the spectrum given by \eqnref{eq:FTSpectrum}, see \figref{Fig.1}a, spanning over $N$ resonance frequencies and with the fine-tuned decay $f(\w) =1/\sqrt{\w^5}$. The strength $m$ of the magnetic moment is normalized such that the energy radiated by the dipole in free space, $U_{\rm rad} \equiv \mu_0m^2 (6\pi^2c^2)^{-1}\int_0^{\infty}\text{d}\omega\phi^2(\omega) \omega^4$, does not depend on $N$. 
Electric and magnetic fields solutions are numerically calculated as a function of space and time. The energy density $u(\rho,z,t)$ is then computed. In \figref{Fig.2}(a,b,c), we show snapshots of these simulations for $N=4,8,12$ modes at similar times. It can be seen that the larger the number of modes, the shorter the duration of the pulses [see \figref{Fig.2}(e)]. The spatial distribution of energy also becomes more focused with increasing $N$, as shown in \figref{Fig.2}(f). In~\cite{SM}, we provide the videos showing the temporal evolution of the energy density for these three cases as well as for a non-tuned spectrum. As discussed above, the mean frequency $\bar \w$ of the dipole barely depends on $N$ and hence, so does $\phi(t)$ [\figref{Fig.2}(d)]. Nevertheless, the high frequency tail of the spectrum has a significant impact on the spatiotemporal features of the generated EM field. We remark that in order to understand the on-axis trajectory of each pulse emitted,  it is very enlightening to solve the similar problem of a dipole placed between two infinite planes made of perfect electric conducting material. By using the method of images, one readily obtains that the trajectory of the $n$th pulse ($n=1,2,\ldots$) is given by the equation $c^2 t^2 = (2n R)^2 + z^2$, which indicates that the group velocity of the $n$th pulse is superluminal after being generated, namely for $ct \gtrsim 2 n R$. This can be neatly corroborated in the videos given in~\cite{SM}.

\begin{figure}[t]
\begin{center}
\includegraphics[width=0.50\textwidth]{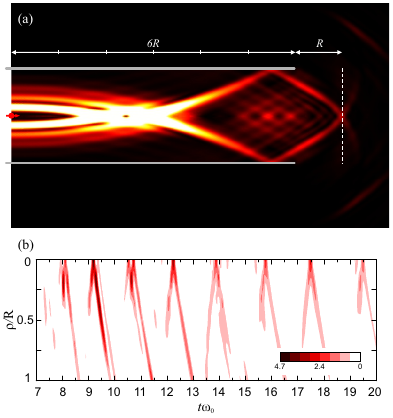}
\caption{(a) Snapshot of normalized EM energy density $u/u_0$ for a finite cylinder with a magnetization spectrum that excites 8 modes [color scale ranges from 0 (black) to 20 (white)] at $t\omega_0\approx12.24$. (b) Energy density evaluated at a distance $R$ from the end of the cylinder [dashed line in panel (a)] as a function of $\rho$ and time.}
\label{Fig.3}
\end{center}
\end{figure}

Motivated by the potential applications, we also numerically study  a cylindrical waveguide of finite length. In particular, we consider a magnetic dipole at the origin of coordinates and a hollow perfect electric conducting cylinder extending from $z\in(-\infty, 6R]$. We analyze the spatiotemporal distribution of the energy density outside the cylinder. As can be seen in \figref{Fig.3}(a), the magnitude of the energy density at distance $R$ outside the cylinder is still comparable to the energy density inside. Furthermore, the main features such as equidistant pulsing and spatial focusing are approximately preserved at a distance $R$ outside the cylinder, see \figref{Fig.3}(b).

Let us now discuss a particular case study. Consider a cylinder of radius $R=2.5\,\text{cm}$ and a magnetic dipole moment generated  with a coil of radius $0.25\,\text{cm}$. The coil is driven by a time-dependent current with a peak intensity of $1\,\text{A}$ and spectrum given by \eqnref{eq:FTSpectrum} with $f(\w)=1/\sqrt{\w^5}$ and $N=8$. The characteristic frequency is then $\omega_0\approx 2\pi\times 1.9\,\text{GHz}$. Inside the cylinder, the temporal duration of an EM pulse is $\delta t\approx8.3\,\text{ps}$, the repetition period $T\approx0.17\,\text{ns}$, and the spot size is given by $\delta\rho\approx0.40\,\text{cm}$. The peak energy density inside the cylinder is $u\approx 5.5\,\mu\text{J}/\text{m}^3$ with a longitudinal magnetic field component given by $B_z\approx 3.7\mu$T. At $2.5\,\text{cm}$ outside the cylinder, the peak energy density is $u\approx 0.19\,\mu\text{J}/\text{m}^3$ and the magnetic field $B_z\approx 0.68\mu$T.

Thorough the Letter we assume that the hollow cylinder consists of a perfect electric conductor. In practice, one should find a material that provides the appropriate boundary conditions over a frequency window that includes several resonance frequencies. Since frequencies depend inversely on the radius of the cylinder, the operating frequency window will strongly depend on $R$. For a cm-sized cylinder, where $\omega_0$ is at the GHz regime, one could use a good conductive metal, such as copper or gold, to obtain the approximate boundary conditions up to frequencies in the terahertz regime~\cite{Pendry2004}. These materials would also introduce some losses, which could be taken into account numerically. Their effect would increase with frequency, ultimately limiting the useful frequency window. Resonant frequencies $\w_n$ also depend linearly on the speed of light $c$. Thus, one could fill the cylinder with a non-structured and non-absorptive medium with a given refractive index to reduce $c$, thereby decreasing the resonance frequencies and effectively increasing the number of modes $N$ that could be perfectly reflected in the waveguide. Another challenge is the fine-tuning of the driven dipole spectrum at high frequencies, namely the precision with which the time-dependent variation of the dipole moment has to be implemented. This precision is then translated into the tiny spatiotemporal features of the emitted EM field. Note that for a sustained generation of ultrafocused EM pulses, one could use a frequency comb to drive the dipole. Interestingly, one could explore if a suitable quantum emitter inside the cylinder can naturally emit radiation with the required optimal tail at high-frequencies, a question we leave for further research. 

In summary, we have discussed an alternative scheme to generate tiny spatiotemporal features of EM fields that is neither based on spatially-structured nor on non-linear EM media. It is solely based on the geometry of a hollow cylindrical waveguide and the fine-tuning of the high-frequency tail of a time-dependent point dipole placed in its interior. The temporal features of the generated EM fields could be used for precise sensing, as similarly done with frequency combs,  and the spatial features for imaging and focusing of either magnetic or electric longitudinal fields.

We acknowledge useful discussions with A.~Rauschenbeutel. This work is supported by the European Research Council (ERC-2013-StG 335489 QSuperMag) and the Austrian Federal Ministry of Science, Research, and Economy (BMWFW).

\clearpage
\widetext
\begin{center}
\textbf{\large Supplementary Material}
\end{center}
\setcounter{equation}{0}
\setcounter{figure}{0}
\setcounter{table}{0}
\setcounter{page}{1}
\makeatletter

Here we solve Maxwell equations for a magnetic and electric dipole inside an infinite hollow cylinder of radius $R$ in the spectral domain. The magnetization and polarization of the dipoles are aligned with the symmetry axis of the cylinder and are denoted by $\boldsymbol{M}\cyl\equiv m\phi(\w)\delta(\rho)\delta(z)/(2\pi\rho)\uv_z$ and $\boldsymbol{P}\cyl\equiv p\phi(\w)\delta(\rho)\delta(z)/(2\pi\rho)\uv_z$. Here $m$ and $p$  denote the strength of the magnetic and electric dipole moment and $\phi(\w)$ denotes a real and symmetric scalar function with dimensions of time. Further, the spectrum is normalized via $\int_0^\infty\text{d}\w\phi(\w)=1$. This gives rise to a charge density $\varrho\cyl=-\nabla\cdot\boldsymbol{P}\cyl$ and a current density $\boldsymbol{j}\cyl=-\im\w\boldsymbol{P}\cyl+\nabla\times\boldsymbol{M}\cyl$. The Maxwell equations thus read
\begin{align}
\nabla\cdot\boldsymbol{E}(\rho,z,\omega)&=-\nabla\cdot\boldsymbol{P}(\rho,z,\omega)/\epsilon_0,\\
\nabla\cdot\boldsymbol{B}(\rho,z,\omega)&=0,\\
\nabla\times\boldsymbol{E}(\rho,z,\omega)&=\im \omega \boldsymbol{B}(\rho,z,\omega),\\
\nabla\times\boldsymbol{B}(\rho,z,\omega)&=-\im\w\boldsymbol{E}(\rr,\omega)/c^2+\mu_0[-\im\omega\boldsymbol{P}(\rho,z,\omega)+\nabla\times \boldsymbol{M}(\rho,z,\w)],
\end{align}
where $\boldsymbol{E}\cyl\text{ and }\boldsymbol{B}\cyl$ denotes the electric and magnetic fields, $\epsilon_0\text{ and }\mu_0$ the vacuum permittivity and permeability and $c$ the speed of light in vacuum. Furthermore we consider the border of the cylinder consisting of a perfect electric conductor, which sets the boundary condition $\uv_\rho\times \EE(R,z,\w)=0$. The electric and magnetic fields can be expressed in terms of the scalar potential $\Phi\cyl$ and the vector potential $\boldsymbol{A}\cyl$ via $\boldsymbol{E}\cyl=-\nabla \Phi\cyl+\im \w \boldsymbol{A}\cyl$ and $\boldsymbol{B}\cyl=\nabla\times\boldsymbol{A}\cyl$. In the Lorenz gauge $\nabla\cdot\boldsymbol{A}\cyl-\im \w  \phi\cyl/c^2=0$ the Maxwell equations lead to the well known scalar and vector wave equations for the potentials, namely $(\Delta+k^2)\Phi\cyl=-\varrho\cyl/\epsilon_0$ and $(\boldsymbol{\Delta}+k^2)\boldsymbol{A}\cyl=-\mu_0\boldsymbol{j}\cyl$, where $k^2\equiv(\w/c)^2$ and $\boldsymbol{\Delta}\equiv\nabla(\nabla\cdot)-\nabla\times(\nabla\times)$. We now proceed to further simplify these equations by expressing the scalar and vector potential in terms of two vector polarization potentials $\boldsymbol{\Pi}_e\cyl$ and $\boldsymbol{\Pi}_m\cyl$,  namely $\Phi\cyl=-\nabla\cdot \boldsymbol{\Pi}_e\cyl$ and $\boldsymbol{A}\cyl=-\im \w \boldsymbol{\Pi}_e\cyl/c^2+\nabla\times \boldsymbol{\Pi}_m\cyl$. 
Note that the Lorentz gauge condition is naturally preserved. Inserting the vector polarization potentials into the corresponding wave equations leads to
\begin{align}
\nabla\cdot&\left[(\boldsymbol{\Delta}+k^2)\boldsymbol{\Pi}_e\cyl+\boldsymbol{P}\cyl/\epsilon_0\right]=0,\\
-\frac{\im\w}{c^2}&\left[(\boldsymbol{\Delta}+k^2)\boldsymbol{\Pi}_e\cyl+\boldsymbol{P}\cyl/\epsilon_0\right]+\nabla\times\left[(\boldsymbol{\Delta}+k^2)\boldsymbol{\Pi}_m+\mu_0\boldsymbol{M}\cyl\right]=0.
\end{align}
Note that the choice of these vector polarization potentials is not unique, since $\boldsymbol{\Pi}'_e\cyl=\boldsymbol{\Pi}_e\cyl+\nabla \times \boldsymbol{f}\cyl$ and $\boldsymbol{\Pi}'_m\cyl=\boldsymbol{\Pi}_m\cyl+\im\w \boldsymbol{f}\cyl/c^2+\nabla g\cyl$ lead to the same vector and scalar potentials for sufficiently well-behaving functions $\ff\cyl \text{ and } g\cyl$. So, after an appropriate gauge transformation, the corresponding vector wave equations for the vector polarization potentials read $(\boldsymbol{\Delta}+k^2)\boldsymbol{\Pi}_e\cyl=-\boldsymbol{P}\cyl/\epsilon_0$ and $(\boldsymbol{\Delta}+k^2)\boldsymbol{\Pi}_m\cyl=-\mu_0\boldsymbol{M}\cyl$. These are two symmetric vector wave equations where the polarization and magnetization themselves are the source terms. For the previously fixed polarization and magnetization we use the ansatz $\boldsymbol{\Pi}_e\cyl=\Pi_e\cyl\uv_z$ and $\boldsymbol{\Pi}_m\cyl=\Pi_m\cyl\uv_z$. One can easily see that both vector polarization potentials are determined by two distinct solutions of the same differential equation, namely
\begin{equation}
\label{eq1}
(\Delta+k^2)\Psi\cyl=-\phi(\w)\delta(\rho)\delta(z)/(2\pi\rho).
\end{equation}
The two solutions, that we denote by $\Psi_\nu\cyl$ with $\nu=0,1$, differ in the boundary conditions $\Psi_\nu\cyl$ has to fulfill. We have that $\Pi_e\cyl=p\Psi_{0}\cyl/\epsilon_0$ with $\Psi_{0}(R,z,\w)=0$ and $\Pi_m\cyl=\mu_0m\Psi_{1}\cyl$ with $\partial_\rho\Psi_{1}(R,z,\w)=0$. The electric and magnetic fields are determined by the same scalar functions. Using the relation between the electric and magnetic fields and the scalar and vector potentials we get outside the region of the dipoles
\begin{align}
\boldsymbol{E}(\rho,z,\w)&=\partial_\rho\partial_z\Pi_e\cyl\uv_\rho-\im \w \partial_\rho\Pi_m\cyl\uv_\varphi+\left(\partial^2_z+k^2\right)\Pi_e\cyl\uv_z,\\
\boldsymbol{B}(\rho,z,\w)&=\partial_\rho\partial_z\Pi_m\cyl\uv_\rho+(\im \w/c^2) \partial_\rho\Pi_e\cyl\uv_\varphi+\left(\partial_z^2+k^2\right)\Pi_m\cyl\uv_z.
\end{align}
Let us now derive the solutions $\Psi_{\nu}\cyl$ of \eqnref{eq1}. The solutions can be expanded in a complete set of orthogonal functions that fulfill the boundary condition. These sets are given by $\lbrace\exp(\im k_z z)\rbrace_{k_z\in\mathds{R}}$ and $\lbrace J_0(j_{\nu n}\rho/R)J^{-2}_{\bar{\nu}}(j_{\nu n})\rbrace_{n\geq 1}$, with $\bar{\nu}\equiv(\nu+1)\text{mod 2}$. Here $J_{\nu}(x)$ denotes the $\nu$-th order Bessel function of the first kind and $j_{\nu n}$ denotes the n-th non-zero root of the $\nu$-th order Bessel function of the first kind. Expanding \eqnref{eq1} in these sets reads
\begin{equation}
(\Delta+k^2)\int_\mathds{R}\text{d}k_z \sum_{n\geq 1} C_{\nu n}(k_z,\w)\frac{J_0(j_{\nu n}\rho/R)}{J^2_{\bar\nu}(j_{\nu n})}\exp(\im k_z z)=-\frac{\phi(\w)}{2\pi^2R^2}\int_{\mathds{R}}\text{d}k_z\exp(\im k_z z)\sum_{n\geq 1}  \frac{J_0(j_{\nu n}\rho/R)}{J^2_{\bar\nu}(j_{\nu n})}.
\end{equation}
Hence, the expansion coefficients $C_{\nu n}(k_z)$ can be easily determined by using the orthogonality relations of the complete sets of functions, which leads to
\begin{equation}
\Psi_\nu\cyl=-\frac{\phi(\w)}{4\pi^2R^2}\sum_{n\geq 1}\frac{J_0(j_{\nu n}\rho/R)}{J^2_{\bar\nu}(j_{\nu n})\sqrt{k^2-(j_{\nu n}/R)^2}}\int_\mathds{R}\text{d}k_z\left[\frac{\exp(\im k_z z)}{k_z+\sqrt{k^2-(j_{\nu n}/R)^2}}-\frac{\exp(\im k_z z)}{k_z-\sqrt{k^2-(j_{\nu n}/R)^2}}\right].
\end{equation}
The integrand is characterized by poles lying on the real (imaginary) axis for $R|k|\geq j_{\nu n}$ ($R|k|<j_{\nu n}$). For $R|k|<j_{\nu n}$ we obtain evanescent fields which will not propagate to the far-field, i.e. the waveguide acts as a high-pass filter with a cut-off frequency $\w^{\text{cut}}_{\nu}=cj_{\nu 1}/R$. Note that \eqnref{eq1} does not distinguish, a priori, between positive and negative valued $k$. Therefore we have to subdivide the problem into four cases $k\lessgtr 0$ and $z\lessgtr 0$. In order to solve the integral we follow the ideas of~\cite{SM:Keller2012}. The integral is undefined until we specify the contour of integration, guided by physical principles. This means that we will shift the poles and choose the integration path such that we retain only outgoing fields. This leads to the following propagating contributions
\begin{equation}
\Psi_\nu\cyl=\frac{\im \phi(\w)}{2\pi R}\sum_{n\geq 1}\Theta(|\tilde\w|-j_{\nu n})\frac{J_0(j_{\nu n}\tilde\rho)}{j_{\nu n}J^2_{\bar\nu}(j_{\nu n})}\frac{\exp\left[\im\, \text{sign}(\w)\sqrt{\tilde{\w}^2-j^2_{\nu n}}|\tilde{z}|\right]}{\text{sign}(\w)\sqrt{(\tilde{\w}/j_{\nu n})^2-1}},
\end{equation}
with the dimensionless frequency given by $\tilde{\w}\equiv\w R/c$ and the dimensionless spatial variables given by $\tilde{\rho}\equiv\rho/R$ and $\tilde{z}\equiv z/R$. The angular and longitudinal components of the electric and magnetic fields can be written as a sum of modes by defining $a_{\nu n}(\rho,\w)\equiv\im\phi(\w)\sqrt{j_{\nu n}^3}J_0(j_{\nu n}\tilde{\rho})/J^2_{\bar\nu}(j_{\nu n})$, $b_{\nu n}=\im \tilde\w (R/j^2_{\nu n})\partial_\rho a_{\nu n}(\rho,\w)$, $\delta_{\nu n}(\w)\equiv\text{sign}(\w)\Theta(|\tilde{\w}|-j_{\nu n})[\sqrt{j_{\nu n}}\sqrt{(\tilde{\w}/j_{\nu n})^2-1}]^{-1}$ and $\alpha_{\nu n}(z,\w)\equiv \text{sign}(\w)\sqrt{\tilde{\w}^2-j_{\nu n}^2}|\tilde{z}|$ and $E_0\equiv p/(2\pi\epsilon_0R^3)$, $B_0\equiv\mu_0 m/(2\pi R^3)$. The corresponding components read
\begin{align}
\frac{E_z\cyl}{E_0}&=\sum_{n\geq 1}a_{0n}(\rho,\w)\delta_{0n}(\w)\exp[\im\alpha_{0n}(z,\w)],\\
\frac{c B_\varphi\cyl}{E_0}&=\sum_{n\geq 1}b_{0n}(\rho,\w)\delta_{0n}(\w)\exp[\im\alpha_{0n}(z,\w)],\\
\frac{B_z\cyl}{B_0}&=\sum_{n\geq 1}a_{1n}(\rho,\w)\delta_{1n}(\w)\exp[\im\alpha_{1n}(z,\w)],\\
\frac{E_\varphi\cyl}{-c B_0}&=\sum_{n\geq 1}b_{1n}(\rho,\w)\delta_{1n}(\w)\exp[\im\alpha_{1n}(z,\w)].
\end{align}

\end{document}